\documentclass[conference]{IEEEtran}
\IEEEoverridecommandlockouts

\usepackage{amsmath}
\usepackage{graphicx}
\usepackage{array}
\usepackage[table]{xcolor}
\usepackage{booktabs}
\usepackage{relsize}

\usepackage{cite}
\usepackage{amsmath,amssymb,amsfonts}
\usepackage{algorithm}
\usepackage{algpseudocode}
\usepackage{caption}
\usepackage{graphicx}

\usepackage{textcomp}
\usepackage{bm}
\usepackage{subcaption}

\usepackage{xcolor}
\def\BibTeX{{\rm B\kern-.05em{\sc i\kern-.025em b}\kern-.08em
    T\kern-.1667em\lower.7ex\hbox{E}\kern-.125emX}}

\begin{document}

\title{\huge Advanced Closed-Loop Method with Limited Feedback for
ISAC}

\author{
\IEEEauthorblockN{
Ashkan Jafari Fesharaki\IEEEauthorrefmark{1},
Yasser Mestrah\IEEEauthorrefmark{2},
Yi Ma\IEEEauthorrefmark{1}, 
Rahim Tafazolli\IEEEauthorrefmark{1}, 
Ibrahim Hemadeh\IEEEauthorrefmark{2},\\
Mohammad Heggo \IEEEauthorrefmark{2},
Arman Shojaeifard\IEEEauthorrefmark{2},
Javier Lorca Hernando\IEEEauthorrefmark{2}
and Alain Mourad\IEEEauthorrefmark{2}
}
\IEEEauthorblockA{\IEEEauthorrefmark{1}6GIC, Institute for Communication Systems, University of Surrey, Guildford, UK, GU2 7XH}
\IEEEauthorblockA{\IEEEauthorrefmark{2} 
InterDigital, London, UK, EC2A 3QR}
}

\maketitle

\begin{abstract}
6G wireless networks are poised to seamlessly integrate communication, computing, localization, and sensing functionalities, ensuring high reliability and trustworthiness. This paper introduces Smart Sensing Feedback (SSF), a limited-feedback framework designed to enhance sensing capabilities while maintaining communication performance. SSF adapts the concept of retransmission from communication to sensing. Specifically, we focus on downlink (DL) bistatic sensing, where the User Equipment (UE) performs measurements from reflected sensing signals and provides feedback to the network (NW). In sensing services, UE reporting can vary significantly due to dynamic factors such as target characteristics, environmental conditions, and UE status. Our results demonstrate that SSF significantly improves sensing quality while preserving communication efficiency. Additionally, it enhances key performance metrics such as probability of detection, latency, and power consumption. These improvements underscore SSF's ability to deliver robust, low-overhead feedback and adaptability to support a wide range of ISAC applications.
\end{abstract}
\section{Introduction}
Future wireless networks will integrate radio sensing capabilities into communication infrastructures, enabling the reuse of hardware and waveforms for both data transfer and environmental sensing, thus significantly improving spectral and energy efficiency \cite{Zhang2021}. This concept, known as integrated sensing and communication (ISAC), has attracted considerable interest in recent 3GPP standardization efforts, particularly in Releases 18–20, due to its potential to enable applications such as unmanned aerial vehicle (UAV) detection \cite{AshDoA, doaa} and tracking, smart transportation through object localization, and cooperative sensing between base stations (BS) and user equipment (UE) for dynamic beam adaptation \cite{3GPP_TR_22}. Executing sensing tasks in bistatic or multi-static modes can face environments with varying dynamics. These dynamics are influenced by slow-varying factors such as path loss, line-of-sight (LOS) conditions, and device impairments, as well as rapidly changing factors like interference, available resources, and target mobility. Consequently, these variations can impact the quality of sensing signal measurements, leading to unnecessary computation and reporting \cite{Zhang2021SignalProcessing}. In addition, deploying dual functionalities necessitates trade‑offs in shared resources such as time, power and bandwidth, since allocating more resources to sensing compromises communication performance, and vice versa \cite{Dong2023}.

Traditional approaches to manage this trade-off often employ isolation strategies either temporally, spatially, or in frequency, to avoid interference \cite{Baig2023}. For instance, time-division multiplexing (TDM) alternates between communication and sensing modes \cite{Smida2023}, while concurrent sensing methods (CSM or hybrid JCAS) partition spatial or frequency resources separately \cite{Wei2022}. Nevertheless, these static strategies reduce adaptability and spectral efficiency: TDM diminishes communication throughput during sensing intervals, while CSM introduces additional complexity such as sophisticated beamforming or full-duplex transceivers \cite{Zhang2021}. Consequently, these drawbacks motivate more dynamic solutions that adaptively balance sensing and communication. A promising direction to enhance adaptability is feedback-driven closed-loop optimization \cite{Dong2023}. Unlike static configurations, ISAC systems with real-time sensing feedback can dynamically adjust transmission parameters to optimize resource allocation \cite{Vinogradova2023}. Recent efforts, such as the extended ARQ (e-ARQ) feedback mechanism introduced in \cite{ericsson}, implement binary hypothesis testing at the receiver to manage sensing tasks. While offering low overhead, these simplistic feedback schemes cannot accurately represent complex sensing conditions, as their fixed decision thresholds lead to inefficient power usage and increased latency upon missed detections. Hence, current feedback approaches exhibit a clear gap which highlights the need for smarter frameworks that dynamically adjust thresholds and provide more informative and adaptive feedback.

To address these shortcomings, this paper proposes a novel limited-feedback framework, called as Smart Sensing Feedback (SSF), for closed-loop beam sweeping in a BS–UE bi-static sensing scenario. Unlike binary feedback methods, SSF employs a multi-hypothesis detection-based feedback protocol with dynamically optimized decision thresholds which enables detailed differentiation between scenarios (e.g., direct target detection, nearby beam presence, broader area occupancy, or absence of targets). Consequently, the network (NW) can optimize transmission tactics in response to dynamic changes and the specific requirements of the sensing task. This includes precise beam steering, power adjustments, modifications to measurement reporting configurations and UE behavior. To enhance adaptability within the feedback framework, we developed an interior-point optimization method to dynamically tune the thresholds in response to varying factors. This approach provides the framework with additional degrees of freedom and greater flexibility. 

Simulation results demonstrate the effectiveness of the proposed SSF framework, achieving approximately 95\% detection probability compared to the 40\% performance of e-ARQ, along with an average reduction in sensing latency by about 2× without additional power usage. Furthermore, utilizing threshold optimization results in a 10\% to 50\% improvement in sensing performance across various power consumption levels.

\section{System Model and Problem Statement}
\subsection{BS-Mobile Bi-static Sensing Architecture}

\begin{figure}[t]
    \centering
    \includegraphics[width=0.45\textwidth]{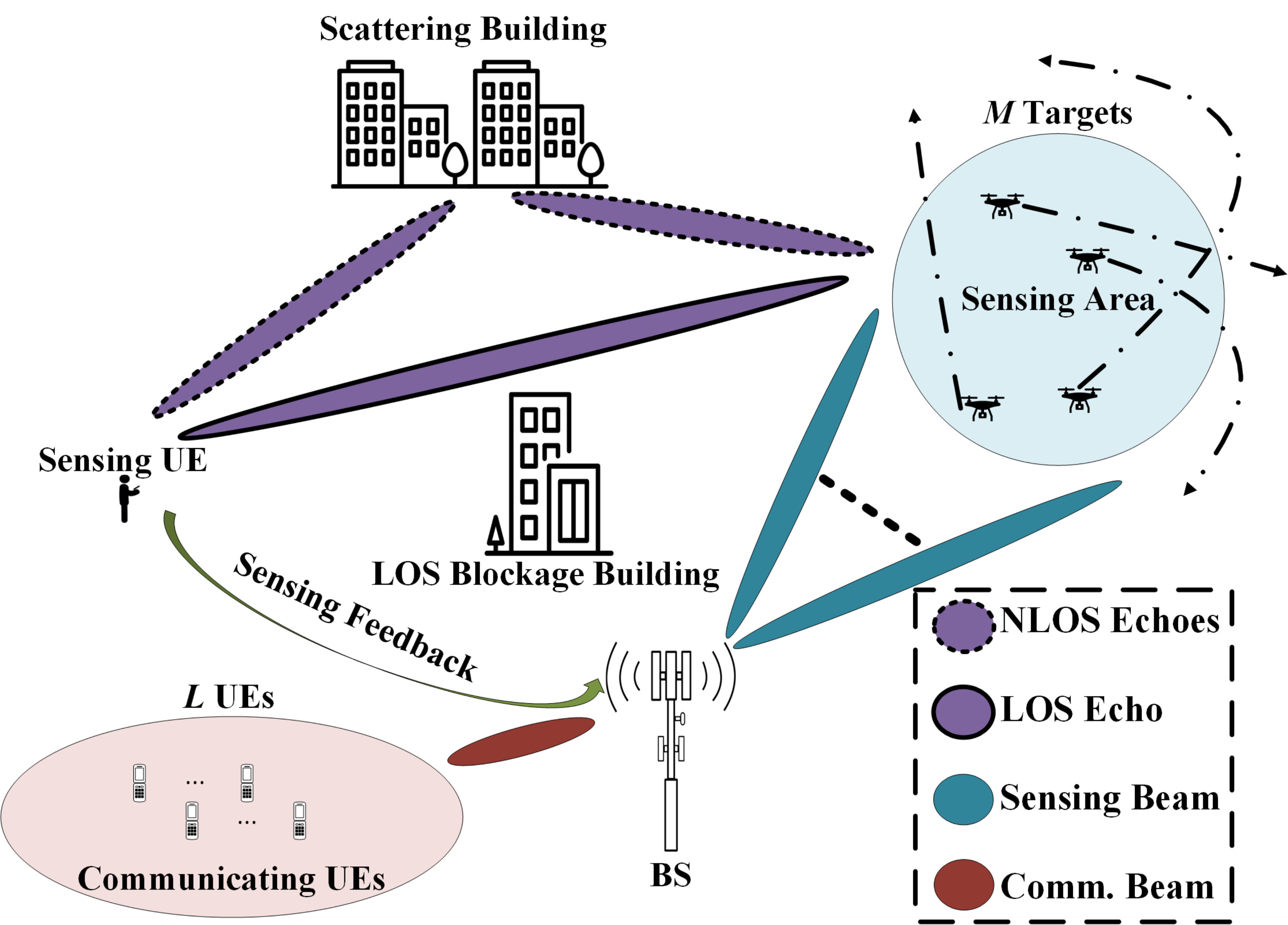}
    \caption{System Geometry: The BS serves $L$ Comm. UEs, while performing bi-static sensing with the Sensing UE.}
    \label{fig:system}
\end{figure}

In this study, an environment is assumed where there is one BS ($T_x$) equipped with a uniform planar array of $N_{\text{BS}}$ antennas, one sensing UE ($R_x$) equipped with a uniform planar array of $N_{\text{UE}}$ antennas, $M_{\text{tg}}$ moving targets and $L$ communicating UEs which are only for communications. A certain sensing area named as $\mathcal{D}_\text{S}$ (divided into $N_{\text{beam}}$ sub-volumes) is defined in this environment where BS is obligated to scan (beam sweep) for its sensing task. On the one hand, $\mathcal{D}_\text{S}$ is designed in a way to make the targets experience both the Line of Sight (LOS) and blockage of LOS (NLOS) toward the sensing UE. On the other hand, the LOS links between the BS and the targets are always available. A scattering object (e.g., walls, buildings, etc.) is assumed to enable the sensing UE to receive the echoes of the targets within the NLOS channel for the time that the LOS link from the targets is blocked. As shown in Fig. \ref{fig:system}, while the targets can have different unique moving trajectories, such as quadratic, linear, and linear with a sudden change in direction, the BS will perform sensing functionality with the sensing UE and communication functionality with the $L$ UEs. 
\subsubsection{Sensing Signal Structure}
The BS uses orthogonal frequency division multiplexing (OFDM) transmission format operating at mmWave with carrier frequency $f_\text{c}$ and subcarrier bandwidth $W_\text{c}$. Moreover, the OFDM subcarriers (with the number of $N_{\text{sub}}$) are uniformly spaced with $W_{\text{sub}}$ (subcarrier spacing). Additionally, $N_{\text{sym}}$ and $T_{\text{sym}}$ denote the number of symbols and their duration, respectively. While the sensing scanning process is synchronized with the duration of the OFDM frame, scanning each sub-volume is assumed to be done for the whole duration of the OFDM frame. This means that it is assumed that the number of symbols transmitted for sensing is equal to $N_{\text{sym}}$. Overall, during the sensing application, while the BS iteratively scans each sub-volume within $\mathcal{D}_\text{S}$ using $N_{\text{sym}}$ OFDM symbols, the reflected echoes from potential targets will be received by the sensing UE.

\subsubsection{Sensing Channel Model}

For the sensing channel model, there are two different channels to be considered: one is the channel from the BS to the targets, which is expected to be a free space path with a line of sight (LOS) link condition. Therefore, the channel between the BS and the $m^{\text{th}}$ target is:
\begin{equation}
    \bm{h}_{\text{BS},q,m} = \sqrt{\beta_{\text{BS}, m}} e^{-j2\pi (q-1) W_{\text{sub}} \tau_{\text{BS}}} \bm{\alpha}(\boldsymbol{\theta}_{\text{BS}, m}) \in \mathbb{C}^{N_{\text{BS}} \times 1}.
    \label{eq:bs-tg}
\end{equation}
where $\beta_{\text{BS}, m} = \frac{\lambda_0^2}{16\pi^2 d_m^2}$ (Friis transmission) and $\tau_{bs}$ denote to path-loss and delay for the LOS link, respectively. The other channel to be considered is the one from the targets to the sensing UE. For simplicity, we considered one LOS and one NLOS link from the targets to the sensing UE. Therefore, the sensing channel would be defined as: 
\begin{equation}
    \begin{aligned}
        \bm{h}_{\text{UE},q,m} =  \overbrace{\sqrt{\beta_{\text{LOS},m}} e^{-j2\pi (q-1) W_{\text{sub}} \tau_{\text{LOS}, m}} \boldsymbol{\alpha}(\boldsymbol{\theta}_{\text{LOS}, m})}^{\bm{h}_{\text{UE}, q, m, LOS}} + \\ \underbrace{\sqrt{\beta_{\text{NLOS}, m}} e^{-j2\pi (q-1) W_{\text{sub}} \tau_{\text{NLOS}, m}} \boldsymbol{\alpha}(\boldsymbol{\theta}_{\text{NLOS}, m})}_{\bm{h}_{\text{UE}, q, m, NLOS}} \in \mathbb{C}^{N_{\text{UE}} \times 1}.
    \end{aligned}  
    \label{eq:tg-ue}
\end{equation}
where $\beta_{LOS}$ and $\beta_{NLOS}$ are the path-loss models as follows while referring $\sigma_m, \sigma_{\text{mp}}$ to the radar cross section (RCS) of the $m^{\text{th}}$ target and the scatter as in the standard bistatic radar equation \cite{Zhang2021SignalProcessing}:
\begin{equation}
\begin{aligned}
\beta_{\text{LOS}, m} = \frac{\sigma_{m}\lambda_0^2}{16\pi^2 d_{\text{tg-UE}}^2} ,
\beta_{\text{NLOS}, m} = \frac{\sigma_{m}\sigma_{\text{mp}}\lambda_0^4}{16^2\pi^4 d_{\text{tg-mp}}^2 d_{\text{UE-mp}}^2}
\end{aligned}
\end{equation}
Same in both channels, $\boldsymbol{\alpha}(\boldsymbol{\theta})$ is defined as the standard far-field steering vector for uniform half-wavelength antenna spacing where $\boldsymbol{\theta} = [\theta_{\text{az}}, \theta_{\text{el}}]$ is the 2-dimensional angle of departure in terms of azimuth and elevation, either from the BS to targets, or targets to the sensing UE, or the scattering object (multi-path) to the sensing UE. 

\subsection{Physical Layer Signaling Model}

\subsubsection{Transmitted Signal}
On each subcarrier, the discrete-time signal emitted by the BS over the $q^{\text{th}}$ subcarrier is
\begin{equation}
    \bm{S}_{q} = \sqrt{P_q} \bm{F} \bm{X}_{q} \in \mathbb{C}^{N_{\text{BS}} \times N_{\text{sym}}},
\end{equation}
for $q = 1, \dots, N_{\text{sub}}$, where $P_q$ is the radiated power, over the $q^{\text{th}}$ subcarrier. Beamforming matrix $\bm{F}$, normalized as $\text{tr}\{\bm{F} \bm{F}^H\} = 1$, is defined as 
\begin{equation}
    \bm{F} = [\gamma_{1} \bm{f}_{1}, \dots, \gamma_{L} \bm{f}_{L}, \gamma_s \bm{f}_s] \in \mathbb{C}^{N_{\text{BS}} \times (L+1)}.
\end{equation}
It will be the collection of the beamforming vectors for the $L$ communicating UEs and one sensing beam, where the coefficients $\gamma$ indicate the power allocation among communicating UEs and sensing. Data symbol matrix $\bm{X}_q$ is defined for storing the transmitted data symbols for the $L$ UEs and the sensing application, over the $q^{\text{th}}$ subcarrier:
\begin{equation}
    \bm{X}_{q} = [\bm{x}_{1,q}^T, \dots, \bm{x}_{L,q}^T, \bm{x}_{s,q}^T] \in \mathbb{C}^{(L+1) \times N_{\text{sym}}}.
\end{equation}
\subsubsection{Received Signal}

In this study, Doppler effects are modeled separately for each link (LOS or NLOS). For instance, $\bm{\psi}_{m, LOS} = [e^{-j2\pi\nu_{m, LOS}1T_{\text{sym}}}, \dots, e^{-j2\pi\nu_{m, LOS} N_{\text{sym}} T_{\text{sym}}}]$ where $\nu_{m, LOS}$ denote the Doppler shift of the $m^\text{th}$ target through the LOS path (for sake of simplicity we assume $\nu_{m, LOS}$ = $\nu_{m, NLOS}$ + $\nu_d$ where $\nu_d$ is a fixed amount). The received signal will be the aggregation of the echoes through both links between the target and the sensing UE. Echoes are the scattered signals which were transmitted by BS and scattered from the targets. Using Eq. \ref{eq:bs-tg} and \ref{eq:tg-ue}, the received signal will be modeled as:
\begin{equation}
\begin{aligned}
 \bm{r}_{q} = \sum_{m = 1}^{M} ( \bm{\psi}_{m, LOS} \odot \bm{w}_{\text{UE}}^H \bm{h}_{\text{UE},q,m,LOS} \bm{h}_{\text{BS},q,m}^{H} \,  \bm{S}_{q} + \\ \bm{\psi}_{m, NLOS} \odot \bm{w}_{\text{UE}}^H \bm{h}_{\text{UE},q,m,NLOS} \bm{h}_{\text{BS},q,m}^{H} \,  \bm{S}_{q}) \, + \bm{z}_q \in & \mathbb{C}^{N_{\text{sym}} \times 1},
 \end{aligned}
\end{equation}
where $\bm{w}_{\text{UE}}$ denotes receive beamformer at the sensing UE, $\bm{z}_q$ represents the noise, and $\odot$ denotes the Hadamard product (i.e., element-wise multiplication between vectors or matrices of the same dimension). Ultimately, by concatenating all $\bm{r}_{q}$ over all the subcarriers we will have $ \bm{r} = [\bm{r}_{1}^T,\bm{r}_{2}^T,\dots,\bm{r}_{N_{sub}}^T]^T \in \mathbb{C}^{N_{\text{sym}}N_{\text{sub}}\times 1} .$

\subsubsection{Reflected Echo Strength Indicator (RESI)}

In this study, without loss of generality we will focus on detection and tracking as an instance of sensing tasks. As discussed in \cite{stat}, detection is about making a decision regarding the presence of a target (or its parameters such as range or Doppler), which leads to a hypothesis test. Therefore, we need to calculate RESI to define each state of the test. For this purpose, the received signal should be filtered via a Delay-Doppler filter in each scan. Similar to  , we assume a grid of Delay ($\tau$) and Doppler ($\nu$) bins, resulting in a grid table with the size of [$N_{\text{del}} \times N_{\text{dop}}$]. The table's values are based on the environment geometry and the targets' velocity. By having this table, the aforementioned filter over subcarrier $q$ in each scan can be introduced as:
\begin{equation}
\begin{aligned}
\bm{g}_{q}(\tau, \nu) = e^{-j 2\pi (q-1)W_{\text{sub}}\tau} \times [e^{-j 2\pi \nu 1 T_{\text{sym}}} , \dots, e^{-j 2\pi \nu n T_{\text{sym}}}].
\end{aligned}
\label{eq:filter}
\end{equation}
Now that the filter is defined ($\bm{g}(\tau, \nu) = [\bm{g}_{1}^T, \dots,\bm{g}_{N_{\text{sub}}}^T]^T \in  \mathbb{C}^{N_{\text{sym}} N_{\text{sub}}  \times 1}$), one can find the specific set of delay and doppler in each scan ($[\tau^{*}, \nu^{*}]$). Accordingly, this specific set, which maximizes the filtered received signal via Eq. \ref{eq:filter} should be selected as in:
\begin{equation}
[\tau^{*}, \nu^{*}] = \arg\max_{\tau, \nu} \, \| \bm{r}^H_{} \bm{g}(\tau, \nu)\|
\label{eq:argmax}.
\end{equation}
With the chosen set of $[\tau^{*}, \nu^{*}]$ in \ref{eq:argmax}, RESI in each scan can be calculated as follows while considering $\sigma_z$ as the standard deviation of the noise \cite{Liu2022}:
\begin{equation}
\text{RESI} = \frac{\| \bm{r}^H \bm{g}(\tau^{*}, \nu^{*})\|}{\sqrt{N_{\text{sub}} N_{\text{sym}} \sigma_z^2}},
\end{equation}
\section{Limited Feedback Design and Optimization}
\subsection{Limited Feedback Design}
Feedback designs, such as Hybrid Automatic Repeat Request (HARQ) in communications, have proven essential for improving the performance while avoiding inefficiency and resource overconsumption (e.g., spectrum) \cite{ARQ}. That being said, a feedback framework is essential in sensing as well, to optimize the trade-off between sensing performance and resource allocation, which affects communications in systems with resource limitations. As discussed in \cite{Vinogradova2023, Baig2023}, with the focus on space-time power distribution, the general ISAC systems' strategies toward power management would be either TDM or CSM. The other resource management considered in this study is beam sweeping (scanning process), which costs spectrum and has a direct impact on the sensing latency. In the following, we propose a novel feedback design for sensing and compare it with the so far other developed feedback designs within ISAC systems. 
\subsubsection{Open-loop Feedback}
In the context of beamforming, open-loop feedback refers to a technique where the transmitter does not receive any feedback from the receiver to adjust its beamforming parameters. Instead, the transmitter relies on pre-determined or estimated channel state information to direct its beams. In the sensing context, a BS will periodically transmit and sweep the sensing signal for a given period. The UE performs the associated measurements and reports the overall measurements to the NW, which processes the results to perform various sensing tasks. This method does not adapt to real-time changes (e.g., channel or target conditions), leading to less optimal performance compared to closed-loop systems that use feedback to continuously adjust the beamforming parameters.
\subsubsection{Extend ARQ Approach}
The authors in \cite{ericsson} developed techniques similar to those used in previous generations' communication modules for sensing functionality, termed extended ARQ (e-ARQ). In e-ARQ, a binary hypothesis test with two thresholds ($\eta_0$ and $\eta_1$) over the RESI divides the system's states into three categories: ACK (reliable detection), NACK (possible target presence), and Lost/Not-found (no target or missed target). However, this simplified division does not fully account for dynamic changes, leading to inefficient feedback actions under certain conditions.
\subsubsection{Smart Sensing Feedback (SSF)}
Without loss of generality, we propose a comprehensive feedback framework based on a multi-hypothesis detector. The multi-hypothesis testing and SSF algorithm design comprises the target state space from the perspective of the sensing UE (e.g., $\mathcal{H}_0, \dots, \mathcal{H}_{K-1}$), the thresholds that define the boundaries for each state ($\eta_0, \dots, \eta_{K-2}$), the conditions for activating or deactivating a given state ($\mathcal{C}_0, \dots, \mathcal{C}_{K-1}$), and the action space associated with each state ($\mathcal{A}_0, \dots, \mathcal{A}_{K-1}$). These conditions can range from measurement conditions (e.g., power, time or Doppler-based), the sensing UE conditions (e.g., velocity, capabilities, mobility), target conditions (e.g., velocity, shape, mobility), to network conditions (e.g., serving cell, available resources). The action space encompasses the behavior of the sensing UE under specific conditions, including actions related to reporting, settings, and measurement behavior. The multi-hypothesis testing framework is designed to be flexible, accommodating a wide range of dynamic factors while meeting diverse sensing requirements. For example, Eq. \ref{eq:multi_hypothesis} employs a multi-hypothesis approach based on four states and conditions derived from power measurements (e.g., RESI). The action state primarily involves performing defined measurements on activated beams by the NW (e.g., via Data Center Interconnect (DCI)) and includes reporting aspects relevant to the target object from the sensing UE's perspective. Reporting from the sensing UE can vary significantly on a state-by-state basis due to dynamic factors such as target and sensing UE's behavior, ranging from binary outputs to full raw Channel Impulse Response (CIR) measurement data.
\begin{equation}
\left\{
\begin{array}{ll}
\mathcal{H}_3, \quad \text{if }  \mathcal{C}_3 := \{\text{RESI} > \eta_2\} \\
\mathsmaller{(\rightarrow M \geq 1 \text{ within the current beam})} \\
\Rightarrow \mathcal{A}_3 := \left\{
\begin{array}{ll} 
\text{Lower the power, Turn off scanning} \\
\text{Utilize communications-favour settings}
\end{array}
\right. \\ \\

\mathcal{H}_2, \quad \text{if } \mathcal{C}_2 := \{\eta_1 < \text{RESI} \leq \eta_2\} \\
\mathsmaller{(\rightarrow M \geq 1 \text{ within the adjacent beams})} \\
\Rightarrow \mathcal{A}_2 := \left\{
\begin{array}{ll} 
\text{Lower the power, Turn off scanning} \\
\text{\small Next Beam $\gets$ Adjacent Beam with higher RESI}
\end{array}
\right. \\ \\

\mathcal{H}_1, \quad \text{if } \mathcal{C}_1 := \{\eta_0 < \text{RESI} \leq \eta_1\} \\
\mathsmaller{(\rightarrow M \geq 1 \text{ present in } \mathcal{D}_{\text{S}})} \\
\Rightarrow \mathcal{A}_1 := \left\{
\begin{array}{ll} 
\text{Lower the power} \\
\text{Next Beam $\gets$ Next adjacent Beam}
\end{array}
\right. \\ \\

\mathcal{H}_0, \quad \text{if } \mathcal{C}_0 := \{\text{RESI} \leq \eta_0\} \\
\mathsmaller{(\rightarrow M = 0)} \\
\Rightarrow \mathcal{A}_0 := \left\{
\begin{array}{ll} 
\text{if {Previously Detected}:} \\ \rightarrow \left\{
\begin{array}{ll} 
\text{\small Higher the power} \\
\text{\small Next Beam $\gets$ Restart}\\
\end{array}
\right. \\ 
\text{else:} \\ \rightarrow \left\{
\begin{array}{ll} 
\text{\small Lower the power} \\
\text{\small Next Beam $\gets$ Next adjacent Beam}\\
\end{array}
\right.
\end{array}
\right.
\end{array}
\right.
\label{eq:multi_hypothesis}
\end{equation}
\subsection{Performance Analysis and Optimization}
As discussed previously, our feedback framework uses a multi-hypothesis test to define\ possible states that the system might encounter, based on the RESI level compared to three thresholds (\ref{eq:multi_hypothesis}). Threshold values can significantly influence sensing outcomes by affecting transmission strategies, power control, and reporting behavior. As a result, assigning customized thresholds for each UE under varying conditions becomes a complex and resource-intensive task. Effective design must account for a range of dynamic factors. This underscores the need for a closed-loop feedback framework that enables intelligent, adaptive threshold selection in response to changing system dynamics.

\subsubsection{MAP-Based Thresholding}
A classical approach to threshold selection in detection theory is through the maximum a posteriori (MAP) criterion \cite{stat}. In a multi-hypothesis detection test with $K$ hypotheses $\{\mathcal{H}_0, \mathcal{H}_1, \dots, \mathcal{H}_{K-1}\}$, each corresponding to a distinct state, the MAP decision rule selects the hypothesis with the highest posterior probability given an observation $x$:
\begin{equation}
    \hat{\mathcal{H}}(x) = \arg\max_i \; p(x | \mathcal{H}_i) P(\mathcal{H}_i)
\end{equation}
Assuming the conditional likelihoods $p(x | \mathcal{H}_i)$ and prior probabilities $P(H_i)$ are known, one can find the decision boundaries $\eta_i$ separating each hypothesis state by solving $ p(x | \mathcal{H}_i) P(\mathcal{H}_i) = p(x | \mathcal{H}_{i+1}) P(\mathcal{H}_{i+1})$ for all state $\mathcal{H}$ pairs.
However, in our context, the MAP framework faces a key limitation: the distribution parameters of RESI under each hypothesis may change during the scenario. Instead, it is shaped by the evolving behavior of the system through feedback. Since the thresholds are the ones that dictate which feedback actions are taken, the resulting RESI distributions become dependent on the threshold values, making the posterior quantities $p(x|\mathcal{H}_i)P(\mathcal{H}_i)$ inherently biased. As a result, MAP optimization becomes ineffective as it often favors one hypothesis and prevents effective threshold adaptation.

\subsubsection{Interior Point-Based Thresholding (IPT)}
To address the nonstationary nature of the problem, we reformulate threshold selection as a black-box optimization problem where the objective is to maximize empirical detection performance. Let $\mathbf{T} = [T_1, T_2, T_3]$ be the decision thresholds which separates four action regions $\{\mathcal{H}_0, \mathcal{H}_1, \mathcal{H}_2, \mathcal{H}_3\}$, in $T_1 > T_2 > T_3$ order. For a given set of thresholds in a scenario, we define successful detection in each time step when $[tg(t)\in Beam(t)] \wedge [\text{RESI}(t) > T_1]$ where $tg(t)$ denotes the target's position. We will then run the feedback-based sensing simulation and estimate the probability of detection as follows:
\begin{equation}
P_{\text{det}}(\mathbf{T}) = \frac{\sum_{t=1}^{T_{\text{S}}}  [tg(t)\in Beam(\mathbf{T},t)] \wedge [\text{RESI}(\mathbf{T},t) > T_1]
}{\sum_{t=1}^{T_{\text{S}}} [tg(t) \in \mathcal{D}_{\text{S}}]}
\label{eq:p-det}
\end{equation}
where $T_\text{S}$ is the number of time steps per simulation. We aim to solve the following constrained optimization problem:
\begin{align}
    \min_{\mathbf{T} = [T_1, T_2, T_3]} \; & f(\mathbf{T}) = -P_{\text{det}}(T_1, T_2, T_3) \\
    \text{subject to } \; & T_3 < T_2 < T_1 \\
                            & \mathbf{T}_{\min} \leq \mathbf{T} \leq \mathbf{T}_{\max}
\end{align}
\begin{figure*}[!t]
    \centering
    \begin{subfigure}[t]{0.325\textwidth}
        \centering
        \includegraphics[width=\linewidth]{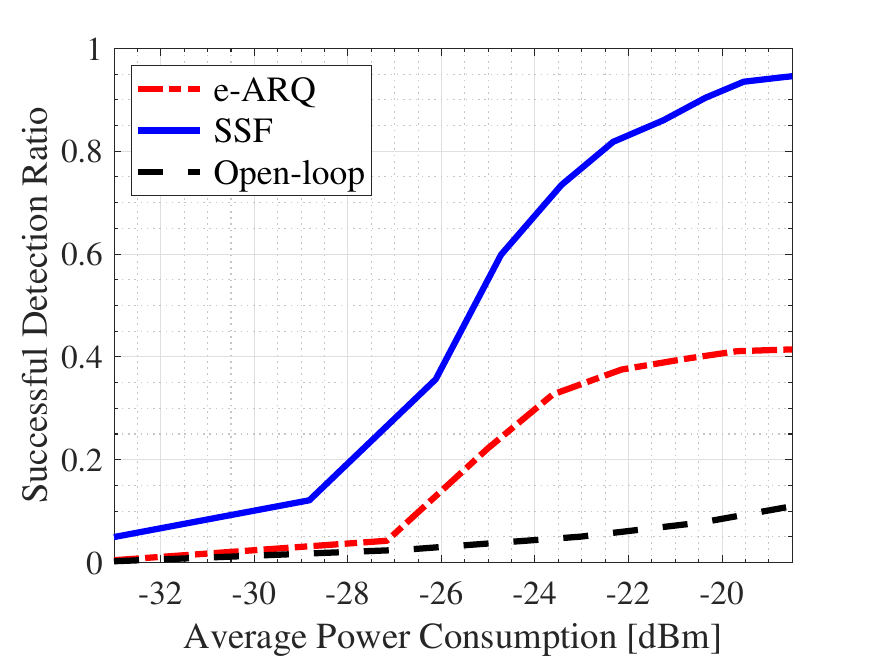}
        \caption{Probability of Detection}
        \label{fig:pd}
    \end{subfigure}
    \hfill
    \begin{subfigure}[t]{0.325\textwidth}
        \centering
        \includegraphics[width=\linewidth]{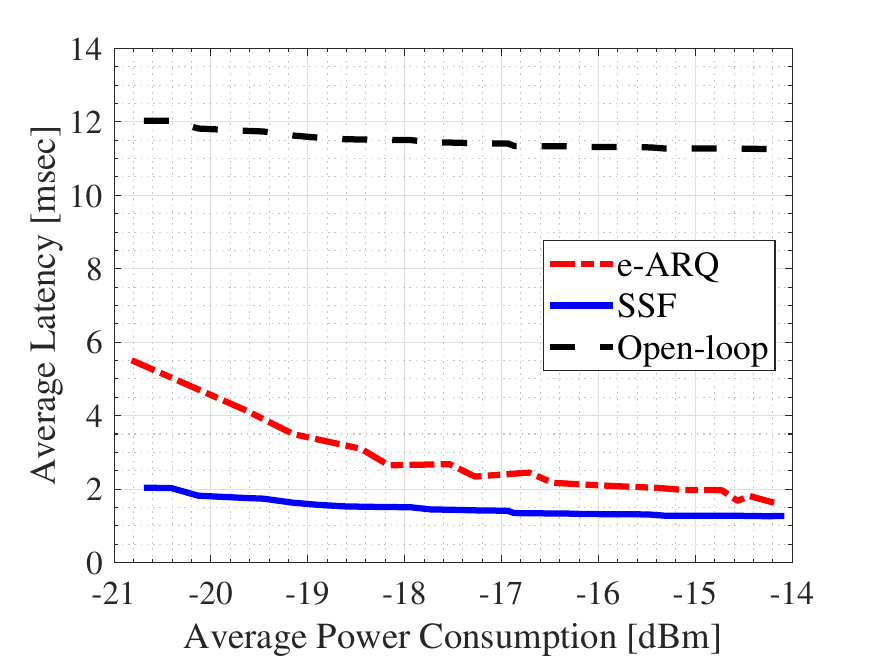}
        \caption{Sensing Latency}
        \label{fig:laten}
    \end{subfigure}
    \hfill
    \begin{subfigure}[t]{0.325\textwidth}
        \centering
        \includegraphics[width=\linewidth]{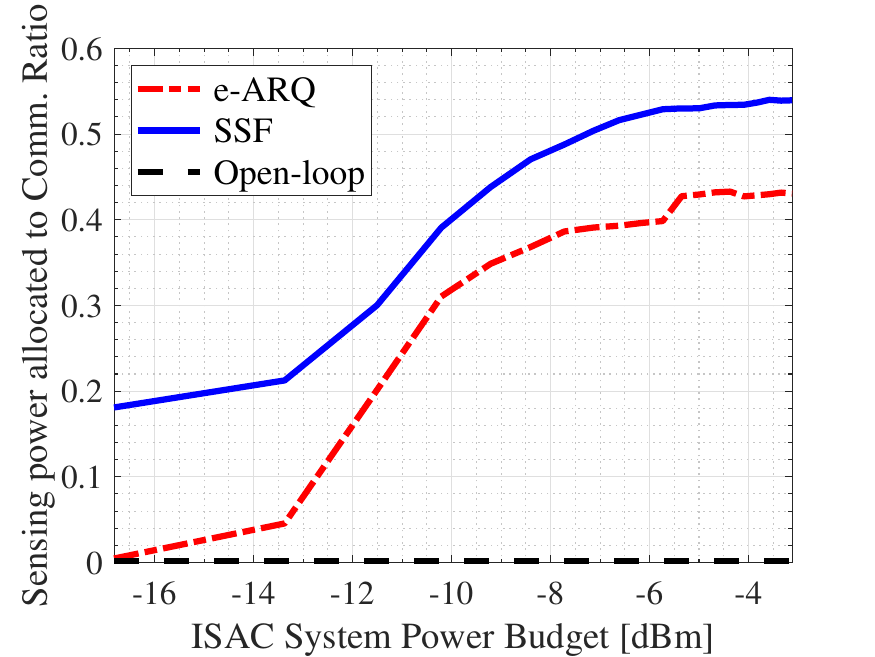}
        \caption{Sensing power allocated to Comm. Ratio}
        \label{fig:ratio}
    \end{subfigure}
    
    \caption{Performance, Latency and Power efficiency comparison between SSF, e-ARQ and Open-loop feedbacks.}
    \label{fig:three-wide}
\end{figure*}
This problem is inherently non-differentiable and lacks an explicit objective function. Therefore, we use a numerical optimization technique based on the interior-point method, a standard approach for constrained nonlinear programming. Starting with initial thresholds from the MAP method, the algorithm estimates the gradient via central finite differences and updates thresholds along the interior-point descent direction. A line search ensures a sufficient decrease in the objective while maintaining feasibility. The solver internally approximates second-order curvature to guide the optimization without explicit computation. Mathematically, the interior-point method solves:
\begin{equation}
\begin{aligned}
    \min_{\mathbf{T} \in \mathbb{R}^3} \; \Phi_\mu(\mathbf{T}) = -P_{\text{det}}(T_1, T_2, T_3) - \\  \mu \left [ \log(T_1 - T_2) + \log(T_2 - T_3) \right]
\end{aligned}
\end{equation}
At iteration $k$, a search direction $\mathbf{d}^{(k)}$ is computed using a quasi-Newton method:
\begin{equation}
    \nabla^2 \Phi_{\mu_k}(\mathbf{T}^{(k)}) \cdot \mathbf{d}^{(k)} = -\nabla \Phi_{\mu_k}(\mathbf{T}^{(k)})
\end{equation}
The solution is then updated as in $\mathbf{T}^{(k+1)} = \mathbf{T}^{(k)} + \alpha^{(k)} \mathbf{d}^{(k)}$.

As shown in Algorithm \ref{alg:threshold_optimization}, this iterative process continues while the barrier parameter $\mu$ is gradually reduced to guide the optimization toward a feasible and optimal solution. 
\begin{algorithm}
\caption{Interior-Point Threshold Optimization}
\label{alg:threshold_optimization}
\begin{algorithmic}[1]
\State Initialize MAP-based thresholds $\mathbf{T}^{(0)} = [T_1^{(0)}, T_2^{(0)}, T_3^{(0)}]$
\State Initialize barrier parameter $\mu_0$ and tolerance $\epsilon$
\For{$k = 1,2,\dots$ until convergence}
    \State Run simulation with $\mathbf{T}^{(k)}$ 
    \State Compute $f(\mathbf{T}^{(k)}) = -P_{\text{det}}(\mathbf{T}^{(k)})$
    \State Construct $\Phi_{\mu}(\mathbf{T^{(k)}}) $
    \State Estimate gradient $\nabla \Phi_{\mu_k}$ using finite differences
    \State Compute descent direction $\mathbf{d}^{(k)}$ from the gradient
    \State Perform line search to determine $\alpha^{(k)}$
    \State Update: $\mathbf{T}^{(k+1)} = \mathbf{T}^{(k)} + \alpha^{(k)} \mathbf{d}^{(k)}$
    \State Decay barrier parameter: $\mu_{k+1} \leftarrow \tau \mu_k$
    \If{objective improvement $< \epsilon$}
        \State \textbf{break}
    \EndIf
\EndFor
\State \Return Optimized thresholds $\mathbf{T}^*$
\end{algorithmic}
\end{algorithm}

This simulation-driven optimization approach is precisely suitable for the feedback dynamics of the system and avoids assumptions about fixed RESI distributions with a moderate complexity cost of $O(nK)$ where $n$ = number of thresholds and $K$ = number of iterations. It results in threshold values that are empirically tuned to maximize the probability of detection under real operating conditions.
\section{Simulation Results and Discussion}
In this section, we aim to evaluate the performance of our proposed feedback design with optimized-thresholds for its multi-hypothesis test in comparison with: 1. Open-loop beam sweeping with no sensing feedback represents the simplest benchmark, meaning that any gain over this scheme directly measures the value of using feedback. 2. e-ARQ \cite{ericsson}, which is the most recent limited-feedback design for bi-static sensing and thus provides a strong state-of-the-art reference. In Table \ref{tab:params}, the simulation parameters are given. For the performance evaluation in terms of sensing, we focused on the probability of detection and average latency. Moreover, to show the superiority of our proposed feedback in terms of communications, we investigated the ratio of sensing power allocation over the whole ISAC system power allocation.
Overall, each scenario would last for 10 seconds ($T_{\text{S}}$) during which $n_{tg}$ targets (here $n_{tg} = 1$) are moving inside and outside of $\mathcal{D}_{\text{S}}$ with a unique trajectory (e.g., quadratic trajectory). For simplicity, we assume that the sensing UE's movement is static.
\renewcommand{\arraystretch}{1}
\rowcolors{1}{gray!15}{white}
\begin{table}[H]
\centering
\caption{Simulation Parameters}
\begin{tabular}{
  >{\centering\arraybackslash}m{1.6cm} 
  >{\centering\arraybackslash}m{1.8cm}| 
  >{\centering\arraybackslash}m{1.8cm} 
  >{\centering\arraybackslash}m{1.8cm}}
\toprule
$[N_{\text{BS}}, N_{\textbf{UE}}]$       & [32, 16]            & BS Beams & 20 
\\ Ant. Spac.           &  $\frac{\lambda_c}{2}$       &  Sweep Range        & $[\frac{\pi}{4}, \frac{3\pi}{4}]$ 
\\ $[f_c ,W_c]$       & [24G, 15k] Hz        & BS Power Rng. & $[-20, -3] \text{ dBm}$ 
\\ Noise Fig.         & 6 dB          & $[T_{\text{sym}}, N_{\text{sym}}]$            & $[100 \mu \text{s}, 100]$ 
\\ $[M_{tg}, v_{tg}]$ & $[1, 3 \text{ ms}^{-1}]$  & 
$[N_{\text{sub}}, W_{\text{sub}}]$ & $[4,10 W_c]$ 
\\ $T_\text{S}$ & 10 sec & $[N_{\text{del}}, N_{\text{dop}}]$ & [10, 10] \\
\bottomrule
\end{tabular}
\label{tab:params}
\end{table}

\subsection{Probability of Detection}
In this study, we refer to the probability of detecting a target when we receive a certain high level of RESI and the BS beam is aligned with the target while the target is within $\mathcal{D}_{\text{S}}$ during each scenario (Eq. \ref{eq:p-det}). As shown in Fig. \ref{fig:pd}, SSF reaches a performance of roughly 95\% compared to e-ARQ's 40\%, over the average sensing power consumption in the BS. This proves SSF needs lower power usage for reliable detection and addresses the use case defined by 3GPP Release 20 report\cite{3GPP_TR_22}. Moreover, it is demonstrated how deploying SSF instead of e-ARQ would prevent saturation in sensing performance.
\subsection{Average Sensing Latency}
This metric will evaluate the average time taken by the system to detect the target when it is present in $\mathcal{D}_{\text{S}}$. As the primary consequence of beam sweeping, lower latency could be interpreted as a more efficient beam sweeping protocol resulting in more efficient spectrum usage. For this purpose, we calculate the average number of times that the systems took for the transition between the state of $\mathcal{H} \neq \mathcal{H}_3$ to $\mathcal{H} = \mathcal{H}_3$. In Fig. \ref{fig:laten}, it is illustrated that while e-ARQ has a lower latency than the Open-loop protocol, SSF has 2 times less sensing latency than e-ARQ, proving SSF's time efficiency in beam sweeping and consequently spectrum usage.
\begin{figure}
    \centering
    \includegraphics[width=0.4\textwidth]{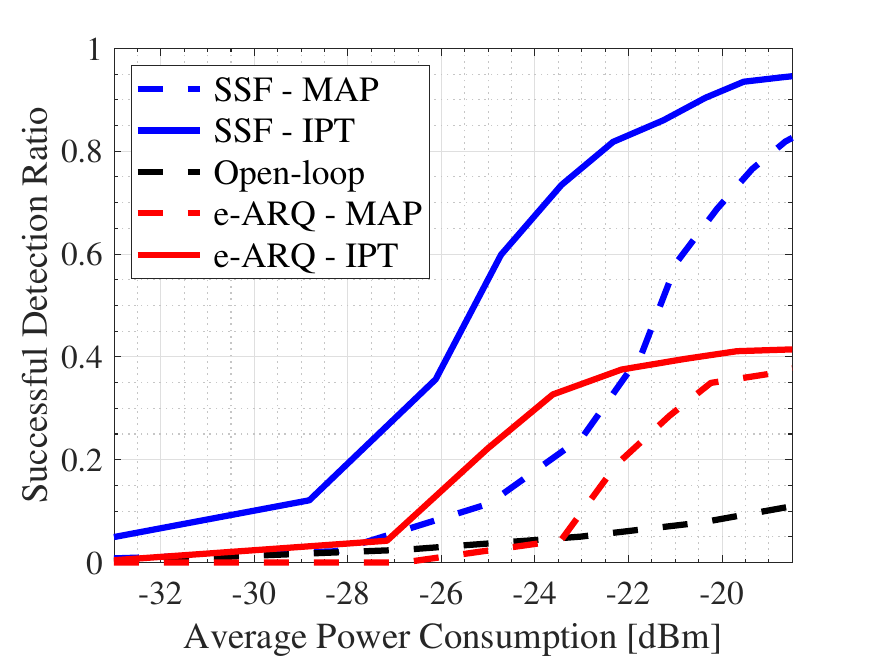}
    \caption{IPT vs MAP thresholding impact on performance.}
    \label{fig:thresh}
\end{figure}
\subsection{Sensing Power Allocation Ratio to Communications}
As this study aims to propose a smart feedback framework for sensing within ISAC systems, it is essential to investigate its impact on communications. For this purpose, we demonstrate the ratio of sensing power reallocated to communications. This reflects how much of the sensing power is redirected to communications, in addition to its initial allocated power. As in Fig. \ref{fig:ratio}, e-ARQ starts to give more power to communications as the system's power budget rises and a lower share of the budget is enough for sensing. However, e-ARQ in contrary to SSF, is not smart enough to give sensing power to communication when the system's whole power budget is too low and detection is nevertheless not feasible. As the whole power budget rises, SSF tries to efficiently allocate power to sensing. Furthermore, it's demonstrated that utilizing SSF instead of e-ARQ would result in an overall 81\% increase in dedicated sensing power to communications.
\subsection{Threshold Optimization}
As previously discussed in section III, hypothesis testing is needed for any feedback design for sensing, which is defined by at least two thresholds (e.g., e-ARQ) or more (e.g., SSF). Fig. \ref{fig:thresh} shows $P_{\text{det}}$ over sensing power consumption for feedback designs using MAP-based thresholding and our proposed simulation-driven optimization thresholding. It is shown that utilizing our proposed threshold optimization (IPT) will result in a performance improvement of 10\% and 50\% in -29 dBm and -24 dBm for the SSF feedback framework. Same happens for the e-ARQ as with 5\% and 25\% improvement. 
\section{Conclusion}
This work introduces Smart Sensing Feedback (SSF), an advanced closed-loop feedback protocol designed to overcome the limitations of static or binary feedback in ISAC systems. By employing a multi-hypothesis detector at the sensing UE and optimizing its thresholds using an interior-point solver, SSF adapts to changing target and channel conditions without significant overhead. Numerical evaluations confirm that SSF consistently achieves significantly higher detection probability, lower sensing latency, and more efficient power re-allocation than both open-loop sweeping and the state-of-the-art e-ARQ scheme, all using the same power budget. These results highlight the practical value of richer, adaptively optimized feedback for balancing sensing reliability with communication throughput in 6G networks. Future studies will extend SSF to dynamically optimize action policy alongside more threshold optimization methods, improving the generalizability.

\bibliographystyle{IEEEtran}  
\bibliography{References}  

\end{document}